\documentclass[superscriptaddress,groupedaddress,nofootnoteinbib,11pt]{article}

\usepackage[lofdepth,lotdepth]{subfig}

\usepackage{jcappub}
\usepackage{bm}
\usepackage{verbatim}

\usepackage{epsf}
\usepackage{graphicx,epsfig}
\usepackage{amsfonts}
\usepackage{amssymb}
\usepackage{xcolor}

\usepackage{soul}

\definecolor{mine}{rgb}{0.2,0.1,0.7}
\definecolor{bb}{rgb}{0.3, 0.5, 1}
\definecolor{bg}{rgb}{0.1, 0.1, 0.5}

\def\half{\frac12}

\def\z{\zeta}

\def\L{\Lambda}

\def\d{\mathrm{d}}

\def\L*{{\cal L}_*}
\def\L{\mathcal{L}}
\def\({\left(}
\def\){\right)}

\def\nn{\nonumber}

\def\<{\langle}
\def\>{\rangle}

\newcommand{\be}{\begin{equation}}
\newcommand{\ee}{\end{equation}}
\newcommand{\bea}{\begin{eqnarray}}
\newcommand{\eea}{\end{eqnarray}}
\newcommand{\ba}{\begin{eqnarray}}
\newcommand{\ea}{\end{eqnarray}}
\newcommand{\beq}{\begin{equation}}
\newcommand{\eeq}{\end{equation}}
\newcommand{\refeq}[1]{(\ref{#1})}
\newcommand{\sdelta}[1]{\!\delta (\mathbf{#1})}
\def\nn{\nonumber}

\def\Tdot#1{{{#1}^{\hbox{.}}}}
\def\z{\zeta}
\def\half{\frac12}
\def\ks{k_{\star}}
\def\d{{\rm d}}
\def\invc{\left(\frac{1}{c_s^2}-1 \right)}
\def\bk{{\bf k}}
\def\fNL{f_{\rm NL}}
\def\la{\langle}
\def\ra{\rangle}
\def\picube{(2\pi)^3}
\def\bkone{\mathbf k_1}
\def\bktwo{\mathbf k_2}
\def\bkthree{\mathbf k_3}

\def\bk{{\bf k}}
\def\bkp{{\bf k'}}
\def\Floc{f_{{\rm NL}}^{{\rm loc}}}
\def\Feq{f_{{\rm NL}}^{{\rm eq}}}
\def\For{f_{{\rm NL}}^{{\rm orth}}}

\begin{document}

\title{Primordial non-Gaussianities after \textit{Planck 2015}: an introductory review}

\author[1,2]{S\'ebastien Renaux-Petel} 
\affiliation[1]{Institut d'Astrophysique de Paris, UMR-7095 du CNRS, Universit\'e Pierre et Marie Curie, 98~bis~bd~Arago, 75014 Paris, France.}
 \affiliation[2]{Sorbonne Universit\'es, Institut Lagrange de Paris, 98 bis bd Arago, 75014 Paris, France}

\vskip 8pt

\abstract{Deviations from Gaussian statistics of the cosmological density fluctuations, so-called primordial non-Gaussianities (NG), are one of the most informative fingerprints of the origin of structures in the universe. Indeed, they can probe physics at energy scales inaccessible to laboratory experiments, and are sensitive to the \textit{interactions} of the field(s) that generated the primordial fluctuations, contrary to the Gaussian linear theory.
As a result, they can discriminate between inflationary models that are otherwise almost indistinguishable. 
In this short review, we explain how to compute the non-Gaussian properties in any inflationary scenario.
We review the theoretical predictions of several important classes of models. We then describe the ways NG can be probed observationally, 
and we highlight the recent constraints from the \textit{Planck} mission, as well as their implications. We finally identify well motivated theoretical targets for future experiments and discuss observational prospects.

\hspace{2cm}

Les d\'eviations \`a la gaussianit\'e des fluctuations cosmologiques de densit\'e, ou non-gaussianit\'es primordiales, nous fournissent des indices pr\'ecieux quant \`a
l'origine des grandes structures de l'univers. Elles permettent en effet de sonder la physique \`a des \'echelles d'\'energie inaccessibles en laboratoire, et sont sensibles aux \textit{interactions} du (ou des) champ(s) \`a l'origine des fluctuations primordiales, contrairement \`a la th\'eorie lin\'eaire gaussienne. Elles nous permettent ainsi de diff\'erencier des mod\`eles autrement presque indistinguables. Dans cette courte revue, nous expliquons comment calculer les propri\'et\'es non-gaussiennes des fluctuations g\'en\'er\'ees pendant tout sc\'enario d'inflation. Nous passons en revue les diff\'erentes pr\'edictions th\'eoriques de plusieurs grandes classes de mod\`eles. Nous d\'ecrivons ensuite la fa\c con dont les non-gaussianit\'es peuvent \^etre contraintes observationnellement et nous soulignons les contraintes r\'ecentes apport\'ees par la mission \textit{Planck}, ainsi que leurs implications. Nous discutons enfin les perspectives observationnelles en identifiant des objectifs r\'ealistes et motiv\'es th\'eoriquement.

\hspace{2cm} 

Published in the French ``Comptes Rendus de l'Acad\'emie des Sciences'' on Inflation.}

\maketitle

\section{Introduction}

Thanks to unprecedented observational efforts in the last two decades, we now have at hand high quality data of the two main cosmological probes, namely the Cosmic Microwave Background (CMB) anisotropies and the Large Scale Structures (LSS). The picture of the primordial universe that emerges from these data is surprisingly simple: all current observations are consistent with the $\Lambda$-Cold Dark Matter model, with initial conditions provided by the simplest inflationary models. The only ingredient of the latter is a single canonical scalar field minimally coupled to gravity and evolving on top of a very flat potential (we call them single-field slow-roll in the following). These scenarios provide a very good fit to the data (see the contribution by Martin, Ringeval and Vennin to this volume) while alternatives to the inflationary paradigm are less compelling (see the contribution by Peter and Lilley). However, despite being phenomenologically successful, these models can not ultimately be considered
as satisfactory, for at least two reasons: they are decoupled from the rest of physics, and they lack a ultraviolet completion. \\

In particular, as soon as one wants to embed the inflationary paradigm into quantum field theory, it becomes surprisingly challenging to realise slow-roll models in an honest-to-God way \cite{Baumann:2014nda} (see also the contribution by Silverstein to this volume). One is then led to consider more complicated inflationary models, involving for instance several degrees of freedom, non-canonical actions or features in the potential. While these models can easily be made degenerate with the simplest ones at the leading-order approximation, their complexities often reveal themselves at next-to-leading-order in the form of so-called \textit{primordial non-Gaussianities}: deviations from Gaussian statistics of the primordial density fluctuations. Cosmological data now put stringent bounds on them, constraining inflationary models in a way that would be otherwise impossible. Here we explain how to compute the non-Gaussian properties of inflationary models and we review the associated theoretical predictions in several important classes of models. We mention the ways non-Gaussianities can be probed observationaly, as well as the current and expected future constraints. This field of research has been very active in the past decade and we can not pay entire justice to it in this short introductory review. For complementary details, we refer the interested reader to the reviews~\cite{Liguori:2010hx,Koyama:2010xj,Chen:2010xka,Byrnes:2010em,Komatsu:2010hc,Wands:2010af,Yadav:2010fz,Desjacques:2010nn,Barnaby:2010sq,Wang:2013zva,Takahashi:2014bxa}, and to the references to the original literature  therein. We use units in which $c=\hbar=M_{{\rm Pl}}=1$.

  \section{Inflation and the origin of the large scale structure of the universe} 
  
\begin{figure}
    \centering
\includegraphics*[width=11.5cm]{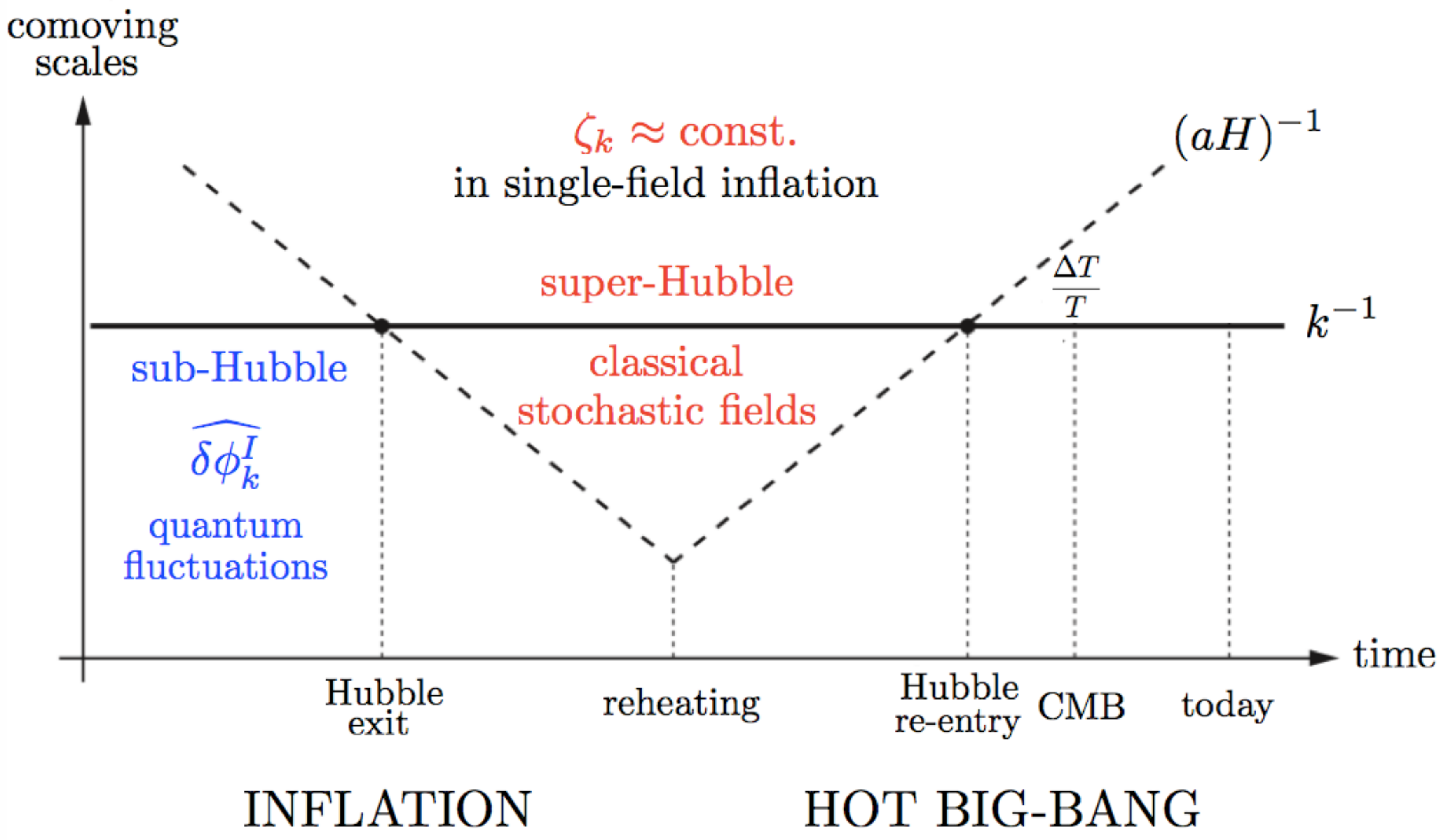}
\caption{The inflationary expansion turns sub-Hubble quantum fluctuations into classical super-Hubble perturbations, whose effects we observe in the CMB and the LSS. \label{f1}}
\end{figure}

 \noindent {\bf From quantum fluctuations to primordial perturbations.}--- Before discussing primordial non-Gaussianites (NG), it is appropriate to remind the reader of some basic facts about the inflationary origin of the large scale structure of the universe, referring her/him to the other contributions in this volume for a more detailed treatment. On cosmological scales, the geometry of the universe can be described at first approximation by a homogeneous and isotropic metric of the (flat) Friedmann-Lema\^itre-Robertson-Walker (FLRW) form:
\be
\overline{\d s}^2=-\d t^2+a(t)^2 \delta_{ij} \d x^i \d x^j
\ee
where $t$ is the cosmic time, $a$ denotes the scale factor of the universe, and $x^i$ are spatial comoving coordinates. The large scale structure of the universe is then described by small fluctuations above this background, which we label by their fixed 3d Fourier wavevectors ${\boldsymbol k}$ in comoving space. The conceptual problems of the hot Big-Bang model all derive from the increase of the comoving Hubble radius $(aH)^{-1}$, where $H=\dot a/a$ denotes the Hubble scale. Inflation solves them by providing an earlier phase in which $(aH)^{-1}$ decreases sufficiently that the observable universe was initially inside the Hubble radius (see Fig.~\ref{f1} and the contribution by Ellis and Uzan to this volume about the causal structure of inflation).
This way, scales of cosmological interest have their physical wavelength
$\lambda=\frac{a}{k=| {\boldsymbol k} |}$
 less than $H^{-1}$ at the beginning of inflation,
exit the Hubble radius during inflation before re-entering it during the Big-Bang era. In this process, unavoidable sub-Hubble quantum fluctuations $\widehat{\delta \phi^I_k}$ of the field(s) driving inflation become classical after Hubble exit \cite{Polarski:1995jg}, 
manifesting themselves in the Big-Bang eras as stochastic fluctuations of the number density $n_X$ of the various components of the universe: radiation, neutrinos, cold dark matter, baryons ... 
Remarkably, all curent data indicate that there is no spatial fluctuation of the composition of the primordial plasma on cosmological scales, \textit{i.e.} $\delta \left(\frac{n_X}{n_Y} \right)=0$ for all $X,Y$. This fact, referred to as the adiabaticity of the primordial fluctuations, implies that the latter can 
be described by one only quantity: the so-called curvature perturbation on uniform density hypersurfaces $\zeta$ deep in the radiation era. It also points to a common origin of these fluctuations, like in the simplest models of inflation driven by a single scalar field. In these scenarios, $\zeta$ becomes constant soon after Hubble exit during inflation, which enables one to make predictions at a much later time (and lower energy) in the radiation era without having to deal with the complicated dynamics of the intermediate phase of reheating after inflation. As we will see, the super-Hubble conservation of the curvature perturbation is lost in general when more than one degree of freedom is active during inflation, like in multifield inflationary models. One should then keep track of cosmological fluctuations on super-Hubble scales
until an adiabatic limit is reached, which requires more attention.\\ 

 \noindent {\bf Background dynamics of inflation.}--- An expanding universe with shrinking comoving Hubble radius is accelerating: $\ddot a >0$. In such a phase, the dimensionless parameter $\epsilon = -\dot H/H^2$ is less than $1$. For simplicity, and in agreement with cosmological data, one assumes in the following that the inflationary expansion is quasi-exponential, \textit{i.e.} that the Hubble parameter varies sufficiently slowly that $\epsilon \ll 1$ and
 $\eta=\dot \epsilon /(H \epsilon) \ll 1$.
More generally, we assume except otherwise stated that all quantities during inflation evolve much less rapidly than the scale factor: $\dot X/X \ll \dot a/a =H$. At leading-order in this slow-varying approximation, $H \simeq {\rm cst}$ and $a(t) \propto e^{H t}$. In terms of the conformal time $\tau$ defined by $\d t =a\, \d \tau$, this gives $a(\tau) \simeq -1/(H \tau)$, where $\tau$ grows from $-\infty$ to $0$ during inflation. Additionally, in terms of the so-called number of e-folds $N$ defined such that $\d N= H \,\d t$, or equivalently $a \propto e^N$, the resolution of the conceptual problems of the Big-Bang model requires that $N \gtrsim 60$.
At the level of classical field theory at least, it is straightforward to achieve a phase with the aforementioned properties by using a scalar field rolling down a sufficiently flat potential. However, there are many other proposals to achieve such a phase and in the following, we show how studying perturbations on top of this background helps to differentiate amongst competing models.

  \section{Primordial power spectrum and higher-order correlators}

Cosmological observations enable one to constrain the large-scale statistical properties of the curvature perturbation. Its two-point correlation function in Fourier space defines the primordial power spectrum $P_\zeta(k)$ such that
 \begin{equation}
\langle\zeta_{{\mathbf k_1}}\,\zeta_{{\mathbf k_2}} \rangle \equiv  \picube\, \sdelta{{\mathbf k_1}+{\mathbf k_2}} \,P_\zeta(k) \,
\label{2-point}
\end{equation}
where $k=|{\mathbf k_1}|=|{\mathbf k_2}|$. On cosmological scales $0.008 \, {\rm Mpc^{-1}} \lesssim k \lesssim   0.1\, {\rm Mpc^{-1}} $, it can be accurately described by a power law
\beq
\frac{k^3}{2 \pi^2}P_\z(k) = A_{\rm s}(\ks) \left(\frac{k}{\ks}\right)^{n_{\rm s}-1}\, 
\label{dvlpt-PR}
\eeq
with the pivot scale $\ks=0.05\, {\rm Mpc}^{-1}$ and the constraint \cite{Ade:2015lrj}
\be
n_s=0.968 \pm 0.006 \quad (68\% {\rm CL}). \label{ns}
\ee
This convincing detection of a percent-level deviation from perfect scale invariance of the primordial power spectrum ($n_s=1$) is one of the most important recent achievement in observational cosmology. It agrees with the predictions of the simplest model of inflation and, together with constraints on the amplitude of primordial gravitational waves, it enables one to pin down the range of inflationary scenarios responsible for the generation of these fluctuations (see the contribution by Martin, Ringeval and Vennin to this volume). 

In addition to this, one can extract complementary information on the \textit{interactions} of the field(s) driving inflation, through the use of higher-order connected correlation functions of the curvature perturbation --- which would vanish for perfect Gaussian statistics --- like the bispectrum
\begin{equation}
\langle\zeta_{{\mathbf k_1}}\,\zeta_{{\mathbf k_2}}\, \zeta_{{\mathbf k_3}}\rangle =  \picube\, \sdelta{\sum_i \bk_i} \,B_\zeta({\mathbf k_i}) \,  \label{3-point}
\end{equation}
and the trispectrum
\be
\la \zeta_{{\mathbf k_1}}\,\zeta_{{\mathbf k_2}}\, \zeta_{{\mathbf k_3}} \zeta_{{\mathbf
k_4}} \ra_c \equiv \picube\, \sdelta{ \sum_i \bk_i}  \, T_\zeta(\bk_i)
\,.
\label{4-point}
\ee
The factors $\,\sdelta{ \sum_i \bk_i}$ above result from statistical homogeneity and imply that the wavevectors $\bk_i$ respectively sustain a triangle and a tetrahedron in Fourier space. Additionally, statistical isotropy entails that the orientations of these polyhedra are irrelevant, so that only their shapes and overall scales matter. This still leaves $3$ independent variables to describe the bispectrum (and $6$ for the trispectrum), which highlights the richness of information contained in these correlation functions compared to the power spectrum \refeq{2-point}, that depends on only one variable. \\

Now concentrating on the bispectrum, it is customary to define its profile $S$ by
\bea
B_\zeta( k_1,k_2,k_3) \equiv (2\pi)^4  \frac{S(k_1,k_2,k_3)}{(k_1 k_2 k_3)^2} A_s^2
\label{def-S}
\eea
where $A_s$ denotes the amplitude of the curvature power spectrum (see Eq.~\refeq{dvlpt-PR}). The overall magnitude of the dimensionless function $S$, which is denoted as the non-linearity parameter $\fNL$, provides an estimate of the importance of the bispectrum. The simplest models of inflation generate a tiny amount of non-Gaussianities \cite{Gangui:1993tt,Acquaviva:2002ud,Maldacena:2002vr}, with $\fNL ={\cal O}(\epsilon,\eta) \ll 1$, under the following hypotheses: a single field drives inflation and generates the primordial fluctuations; it has a standard kinetic term and a smooth potential; its perturbations are in a vacuum state. However, as we will see, violating any one of these restrictive conditions can result in observably large NG, $ | \fNL | \gg 1 $, while recent constraints give the interesting bound $\fNL \lesssim {\cal O}(10)$.

Besides this qualitative estimate, the detailed geometrical dependence of $S$ on the configuration of the wavevectors enables one to differentiate amongst the different classes of inflationary models beyond single-field slow-roll \cite{Babich:2004gb}. For example, scenarios with \textit{multiple} light degrees of freedom generate a bispectrum than can be large in squeezed configurations, \textit{i.e.} for $k_3 \ll k_1 \simeq k_2$, in contrast with single-field models, which predict a vanishingly small bispectrum in this limit. Additionally, models with important derivative interactions, for instance with non-canonical kinetic terms, are characterized by a bispectrum that is maximum for wavevectors of similar magnitude, \textit{i.e.} around equilateral configurations $k_1 \simeq k_2 \simeq k_3$. Scenarios with features also come hand with hand with their signatures in the form of specific oscillatory bispectra. This way, cosmologists have identified in the past decade a useful dictionary between several broad classes of inflationary mechanisms for the generation of primordial fluctuations and the associated type of non-Gaussianities. In this respect, it is useful to decompose the geometrical dependence of $S$ in two parts: its shape and its scale-dependence. In the simplest `smooth' models of inflation, without features, modulations of the potential or excited states for instance, $S(k_1,k_2,k_3)$ depends mainly on the ratios between the norms of the wavectors, \textit{e.g.} $k_2/k_1$ and $k_3/k_1$, and mildly on their overall scales. In the following, we order the $k_i$'s such that $k_3 \le k_2 \le k_1$, so that this shape information can be represented by plotting the two-dimensional function $S(1,x_2,x_3)$, where $0 \le x_2,x_3 \le 1$ and $1 \le x_2+x_3$ (to satisfy the triangle inequality). In more complicated, non-scale invariant, models, this has to be complemented by the study of the dependence of $S$ on the overall scale $K=k_1+k_2+k_3$ while keeping the ratios $k_2/k_1$ and $k_3/k_1$ fixed.

\section{Methods to calculate primordial non-Gaussianities}
\label{Methods}

In this section, we explain how to calculate higher-order correlation functions in any inflationary scenario. For definiteness, we consider models involving scalar fields and a cosmological metric that takes the flat FLRW form at the background level. However, the methods we present are generic and are easily applicable to models with other matter contents, such as vector fields, and other background symmetries.\\

\noindent {\bf Cosmological perturbation theory.}--- We start by decomposing the metric and the fields $\phi^I$ $(I=1, ..., M)$ into their background and fluctuating parts:
\bea
g_{\mu \nu}(t,x^i)&=&\bar g_{\mu \nu}(t)+ \delta g_{\mu \nu}(t,x^i) \\
\phi^I(t,x^i) &=& \overline{\phi^I}(t)+ \delta \phi^I(t,x^i)
\eea
where the background parts only depend on cosmic time. We then expand the (gravitational and matter) action about the background solutions of the equations of motion, which we write schematically as:
\bea
S= \bar S+S^{(2)}(\delta g_{\mu \nu},  \delta \phi^I) +\underbrace{S^{(3)}(\delta g_{\mu \nu},  \delta \phi^I)+S^{(4)}(\delta g_{\mu \nu},  \delta \phi^I)+\ldots }_{S_{{\rm interactions}}} \label{S-perturbative} 
\eea
The second-order part governs the behaviour of linearised fluctuations, while the higher-order parts encode their interactions. A crucial subtlety in these calculations comes from the ambiguity in identifying spacetime points between the idealised background and the true perturbed universe, and the gauge freedom it entails \cite{Bardeen,Malik:2008im}. As a consequence, one must pay special attention to the proper identification of the true dynamical degrees of freedom and the construction of the corresponding gauge-invariant variables. For instance, a model with canonical Einstein-Hilbert action and $N$ scalar fields contains $4+N$ (Lagrangian) degrees of freedom: two tensor modes, so-called gravitational waves, two tensor modes, and $N$ scalar modes. These different sectors are decoupled at the linearised level and evolve very differently. Tensor modes are only sensitive to the background spacetime geometry, \textit{i.e.} to the evolution of the scale factor. Without further sources, vector modes decay due to the expansion of the universe. On the contrary, details of the evolution of the scalar perturbations depend on the model at hand. They provide the main sources of the curvature perturbation $\zeta$, which imprinted the CMB anisotropies and seeded the formation of LSS. We concentrate on them in the following, and discard the other types of fluctuations for simplicity of presentation. \\

\noindent {\bf Quantisation.}--- Once the second-order action is known, one can proceed to its canonical quantisation by promoting the propagating degrees of freedom to quantum operators. The procedure is well known and is straightforward in the single-field case \cite{Mukhanov-Feldman-Brandenberger} but it should be taken care of carefully in multifield scenarios (see \textit{e.g.} Ref.~\cite{Weinberg-cosmology}). The normalisation of fluctuations, which is arbitrary in a classical context, is fixed upon imposing the canonical commutation relations. Eventually, a choice of vacuum $|0 \rangle$ should also be made: deep inside the Hubble radius ($k \gg a H$), fluctuations are not sensitive to the expansion of the universe and, following the equivalence principle, it is legitimate to favor the approximate Minkowski vacuum, also called Bunch-Davies vacuum \cite{Bunch-Davies} (see below for a discussion of the impact of an excited initial state). This completes the quantisation procedure, leaving only the task of accordingly solving the linear equations of motion deduced from the second-order action, either analytically or numerically. Finally identifying the vacuum expectation value of observables in the quantum theory with the statistical ensemble average of the corresponding variables in the classical theory, one can then extract predictions, for the curvature power spectrum \refeq{2-point} in particular.  \\ 

\noindent {\bf \textit{In-in} formalism.}--- After having quantified the linear Gaussian theory, and identified the interacting action in Eq.~\refeq{S-perturbative}, one can determine higher-order correlation functions using the so-called \textit{in-in} (also called Schwinger-Keldysh) formalism \cite{Jordan,Calzetta-et-al,Weinberg:2005vy}. Starting from first principles in quantum field theory, it shows that the expectation value of an observable $O(t)$ can be computed perturbatively as
\bea
 \langle in |  O(t) | in \rangle = \langle 0| \left[ \bar T \exp \left( i \int_{-\infty(1-i \epsilon)}^t H_I(t') \d t' \right) \right] O^I(t) \left[ T \exp \left( -i \int_{-\infty(1+i \epsilon)}^t H_I(t'') \d t'' \right)\right] |0\rangle 
 \nn
 \label{equ:in-in}
\eea
where $ | in \rangle$ is the vacuum of the interacting theory at some moment $t_i$ in the far past, $T$ denotes the time-ordered product, the $I$'s indicate the use of the interaction picture and $H_I$ is the interacting Hamiltonian.
At first-order in the latter, as relevant for the calculation of the tree-level bispectrum, one finds
\beq
\label{equ:first}
\langle O(t) \rangle^{(1)}= 2 \,{\rm Re} \left[ - i \int_{-\infty}^t \d t' \langle 0| O^I(t) H_{I}(t') | 0 \rangle \right]\, ,
\eeq
whereas this reads 
\bea
\label{equ:second}
\langle O(t) \rangle^{(2)} &=& \int_{-\infty}^t \d t' \int_{-\infty}^t \d t'' \langle 0 | H_{I}(t') O^I(t) H_{I}(t'') | 0 \rangle \nonumber \\
&-& 2\, {\rm Re} \left[ \int_{-\infty}^t \d t'   \int_{-\infty}^{t'} \d t''  \langle 0| O^I(t) H_{I}(t') H_{I}(t'') | 0 \rangle \right]\, \nonumber
\eea
at second order, which is necessary for computing the trispectrum for instance (see Ref.~\cite{Chen:2009zp} for general expressions at higher orders). These terms are then simply evaluated by applying Wick's theorem (let us recall that all fields here are in the interaction scheme and hence are free fields). Eventually, note that, as usual in quantum field theory \cite{Peskin-Schroeder,Weinberg-QFT}, as one wishes to compute these expectation values in the vacuum of the interacting theory, and not in the vacuum of the free (linear) theory, one should slightly deform the integration contours in the complex plane (the standard $i \epsilon$ prescription), which turns off the interactions in the far past and renders the time integrals well-behaved then.\\

\noindent {\bf $\delta N$ formalism.}--- The \textit{in-in} formalism is very general and can be \textit{a priori} applied to any inflationary model. However, in multifield models, it is often difficult to obtain accurate analytical expressions for the linear mode-functions beyond a few e-folds after the time of Hubble crossing. For this reason, it is useful to have at hand another formalism called $\delta N$ \cite{Sasaki:1995aw,Sasaki:1998ug,Lyth:2004gb,Lyth:2005fi,Dias:2012qy}. The latter relies on the separate-universe picture \cite{Wands:2000dp}. It states that each super-Hubble patch --- as relevant for dealing with fluctuations on super Hubble-scales $k \ll a H$ --- evolves like a separate FLRW universe which is locally homogeneous and evolves independently from its neighbours. However, patching these regions together enables one to track the evolution of the curvature perturbation on large scales, just by using background quantities. More specifically, one can express the curvature perturbation $\zeta(t,x^i)$ as the difference between the number of e-folds of expansion $N$ between an initial flat arbitrary hypersurface and the uniform energy density hypersurface at time $t$, and its corresponding background value:
\be
\label{deltaN-1}
\zeta(t,x^i)=N(t,t_{\star};x^i)-\bar N(t,t_{\star}) \,.
\ee
This formula is completely generic and non-perturbative. Let us now specify it to attractor solutions of multifield inflationary models\footnote{A generalised $\delta N$ formalism has been developed to deal with more complicated background dynamics \cite{Lee:2005bb}.}. Their dynamics is entirely dictated by specifying the initial values of the scalar fields, \textit{i.e.} the spatial dependence in $N(t,t_{\star};x^i)$ can be encoded as $N(t,\phi^A(t_\star,x^i))$. One can then Taylor-expand Eq.~\refeq{deltaN-1} as a function of the field fluctuations on the initial flat hypersurface.
By writing $\phi^A_\star={\bar \phi^A_\star }+Q^A_\star$,
one obtains
\bea
\label{dev-deltaN}
\zeta&=&N_A Q^A+\half N_{AB} Q^A Q^B +\frac{1}{6} N_{ABC} Q^A Q^B Q^C+\ldots \label{zeta-deltaN}
\eea
where
\be
N_A \equiv  \left. \frac{\partial N}{\partial \phi^A_\star} \right|_{{\bar \phi^A_\star }}  \,, \quad N_{AB} \equiv  \left. \frac{\partial^2 N}{\partial \phi^A_\star \partial \phi^B_\star} \right|_{{\bar \phi^A_\star }} \,\ldots
\ee
From Eq.~\refeq{zeta-deltaN} it is straightforward to deduce the statistical properties of the curvature perturbation in terms of the ones of the scalar fields. Defining their spectra, bispectra and trispectra\footnote{A prime on correlation functions indicate that one omits the ever-present factor $\picube\, \sdelta{ \sum_i \bk_i}$.}:
\bea
\la Q_\bk^A Q_\bkp^B \ra' & \equiv& C^{AB}(k)\, \label{def-CAB} \\
\la Q^A_{{\mathbf k_1}}\,Q^B_{{\mathbf k_2}}\, Q^C_{{\mathbf k_3}}  \ra'&
\equiv& B^{ABC}(\bk_i)\, \label{defbispectrum} \\
 \la Q^A_{{\mathbf k_1}}\,Q^B_{{\mathbf k_2}}\, Q^C_{{\mathbf k_3}}
Q^D_{{\mathbf k_4}} \ra_c' &\equiv & T^{ABCD}( \bk_i)\, 
\label{deftrispectrum}
\end{eqnarray}
one obtains the expressions of the tree-level primordial spectrum, bispectrum and trispectrum as \cite{Allen:2005ye,Byrnes:2006vq}
\be
P_\zeta(k)=N_A N_B C^{AB}(k)
\label{Spectre-general}
\ee
\bea 
&&\hspace*{-2.3em} B_\zeta(k_1,k_2,k_3) = N_A N_B N_C B^{ABC}(k_1,k_2,k_3)+ N_A N_{BC} N_D \left[ C^{AC}(k_1) C^{BD}(k_2) +{\rm 2 \,perms.} \right] 
 \label{zetabispectrum} 
\eea  
and
\bea
&& T_\zeta(\bk_i) = N_AN_BN_CN_D T^{ABCD}(\bk_i)+ N_{A B}N_C N_D N_E \left[ C^{A C}(k_1)B^{BDE}(k_{12},k_3,k_4) \right] \nn \\
&&\hspace*{-2.0em} + N_{A B}N_{C D}N_E N_F \left[C^{BD}(k_{13})C^{AE}(k_3)C^{CF}(k_4) \right] 
+N_{A B C}N_D N_E N_F
\left[C^{A D}(k_2)C^{B E}(k_3)C^{C F}(k_4)\right]\, \nn \\
&&\hspace*{+3.0em} + {\rm permutations}
\label{trispectrum-general}
 \eea
where $k_{ij}=|\bk_i+\bk_j |$ and `permutations' indicate that one should add the respectively 11, 11 and 3 inequivalent permutations of the $\bk_i$ to the last three lines, so that the trispectrum is totally symmetric.\\

\noindent {\bf Interpretation.}--- For definiteness, let us now specify the time of the initial flat hypersurface to be a few efolds after the period of Hubble crossing for the relevant cosmological scales. The \textit{in-in} formalism then enables one to accurately predict the correlation functions of the field perturbations \refeq{def-CAB}-\refeq{deftrispectrum}. The expressions above, although formal, yet deliver a useful information. For instance, one can see that the bispectrum \refeq{zetabispectrum} is a sum of two contributions: the first line, stemming from the linear term $N_A Q^A$ in Eq.~\refeq{dev-deltaN}, results from the linear transfer to the curvature bispectrum of the intrinsic bispectra of the field fluctuations $B^{ABC}$ around Hubble crossing. The second line, proportional to the second-order coefficient $N_{AB}$ in the expansion \refeq{dev-deltaN}, is present even for perfectly Gaussian field fluctuations, and comes from the non-linear relation between the latter and the curvature perturbation $\zeta$. A similar discussion holds for the trispectrum \refeq{trispectrum-general}: the first line simply results from the linear transfer of the field trispectra to the curvature perturbation. The last two lines, on the opposite, solely come from the respectively quadratic and cubic terms in the expansion \refeq{dev-deltaN}. The second contribution, of a type that is not present for the bispectrum, is more complicated, as it mixes the intrinsic bispectra of the field fluctuations and the nonlinear (quadratic) term in the relation \refeq{dev-deltaN}.

 Barring it, one can therefore classify the contributions to the bispectrum and trispectrum in two main types. The first ones originate from purely quantum effects around (and possibly before) the time of Hubble crossing. Their precise geometrical forms depend on the model at hand, and their determinations require the use of the \textit{in-in} formalism. They can be important in either single or multi-field scenarios, in models with large derivative interactions in particular \cite{Alishahiha:2004eh,Chen:2006nt,Langlois:2008wt,Langlois:2008qf}. The second type comes from the classical non-linear relation between the curvature perturbation and the field fluctuations around the time of Hubble crossing. Their shapes are entirely fixed by the two-point correlation functions of the field fluctuations $C^{AB}$, and therefore take universal forms in the simplest models, as we will see. Their importance relies on large non-linear terms in the expansion \refeq{dev-deltaN}, and therefore a large sensitivity of the number of e-folds of expansion to the initial field values. No such sensitivity is present in (attractor) single-field models, so that this type of contribution is characteristic of multi-field models with non-trivial dynamics on super-Hubble scales (see below for details).

\section{General single-field models}
\label{Single-field}
 
In this section, we discuss the non-Gaussian properties of the curvature perturbation generated in single field models of inflation. For simplicity, we restrict ourselves to the representative class of models known as `\textit{k}-inflation' \cite{ArmendarizPicon:1999rj,Garriga:1999vw}. They generalise the inflaton Lagrangian from its canonical form ${\cal L}=X-V(\phi)$, where $X \equiv -\frac12 g^{\mu \nu} \partial_{\mu} \phi \partial_{\nu} \phi$ is the field's kinetic term, to a general function ${\cal L}=P(X,\phi)$. Such models can support a phase of inflation of a very different type from slow-roll scenarios.
For example, the prototypical model of string-inspired Dirac-Born-Infeld (DBI) inflation \cite{Silverstein:2003hf,Alishahiha:2004eh}, with $P=-1/f(\phi) \left(\sqrt{1-2 f(\phi) X}-1 \right)-V(\phi)$ and a positive $f(\phi)$, can achieve inflation on top of very steep potentials, not through Hubble friction like in standard models, but because the noncanonical kinetic term bounds the speed of the inflaton, irrespective of the potential: $2{\bar X}=\dot {\bar \phi}^2 \leq 1/f(\phi)$.

Although it may not seem intuitive at first sight, it is useful in single-field models to choose a coordinate system in which the field $\phi$ has no fluctuation $\delta \phi$. It is always possible to choose such a coordinate system, called the uniform inflaton gauge, and to additionally impose that the spatial part of the metric takes the form (still discarding tensor perturbations)
\be
g_{ij}=a^2(t) e^{2 \zeta(t,x^i)} \delta_{ij}\,.
\label{gij}
\ee
The variable $\zeta$ defined in this way coincides in this gauge with the gauge-invariant curvature perturbation we are interested in. It is therefore quite convenient to directly deal with it in calculations, without having to make explicit its relation with $\delta \phi$ at non-linear orders. By solving the constraint equations to express the non-dynamical parts of the metric in terms of the propagating degree of freedom $\zeta$ (this is most easily done by using the ADM formalism \cite{Arnowitt:1962hi}), one can derive the second-order action
\be
S^{(2)}=\int \d t\, \d^3 x \, a^3 \frac{\epsilon}{c_s^2} \left(\dot \zeta^2-c_s^2 \frac{(\partial_i \zeta)^2}{a^2}   \right)\, \label{S2-single-field}
\ee
where
 \be
 c_s^2=\frac{P_{,X}}{P_{,X}+ 2X P_{,XX}}
 \ee
  is called the (squared) sound speed of perturbations. It equals one in canonical models, in which $P=X-V(\phi)$, but it can be much less than one in general, like in the example of DBI models where $c_s=\sqrt{1- f(\bar \phi)\dot {\bar \phi}^2}$.  
At leading-order in the slow-varying approximation, one can analytically solve the equation of motion deduced from Eq.~\refeq{S2-single-field}:
\be
\zeta_k(\tau)=\frac{iH_*}{\sqrt{4 \epsilon_* c_{s*} k^3}} \left(1+i k c_{s*} \tau \right)e^{-i k c_{s*} \tau}\, \label{mode-function}
\ee
where $*$ denotes evaluation at the time of the sound horizon crossing $k c_s =a H$. As explained above, one has chosen Bunch-Davies initial conditions and the normalisation comes from the quantization procedure. 
It is then straightforward to deduce the primordial power spectrum
\be
\frac{k^3}{2 \pi^2}P_\z(k)=\frac{k^3}{2 \pi^2} \lim_{-k \tau \to 0} |\zeta_k(\tau) |^2= \left( \frac{H^2}{8 \pi^2 \epsilon c_s} \right)_*\,,
\ee
with a mild scale dependence given by the slight dependence of $H, \epsilon$ and $c_s$ on the time of evaluation $*$. \\

\noindent {\bf Equilateral and orthogonal non-Gaussianities.}--- Now moving on the bispectrum, one first calculates the leading-order cubic action \cite{Seery:2005wm,Chen:2006nt}:
\bea
&&\hspace*{-2.0em}S^{(3)}_{{\rm L.O.}}=\int \d t \, \d^3 x\, a^3 \epsilon  \invc \left[ \zeta \frac{(\partial \zeta)^2}{a^2}-\frac{3}{c_s^2} \zeta \dot \zeta^2  \right]
+\int \d t \, \d^3 x\, a^3 \frac{\epsilon}{H c_s^2} \left(\frac{1}{c_s^2}-1 -\frac{2 \lambda}{\Sigma}\right) \dot \zeta^3\,, \label{S3-k-inflation}
\eea
where one concentrated on the main sources of non-Gaussianities,
\bea
\Sigma=X P_{,X}+2 X^2 P_{,XX}=\frac{H^2 \epsilon}{c_s^2}\\
\lambda=X^2 P_{,XX}+\frac23 X^3 P_{,XXX}\,
\eea
and all parameters are taken to be constant in this approximation. Using Eq.~\refeq{equ:first}, one can then deduce the shape of the bispectrum as $S=S^\lambda+S^c$, with 
\be
S^\lambda= \left(\frac{1}{c_s^2}-1
- \frac{2\lambda}{\Sigma} \right)
\frac{3k_1k_2k_3}{2K^3}
\label{Alam}
\ee
and 
\bea
&&S^c=\left(\frac{1}{c_s^2}-1\right)  \frac{1}{k_1k_2k_3}
\left[-\frac{1}{K}\sum_{i>j}k_i^2k_j^2+ \frac{1}{2K^2}
\sum_{i\neq j}k_i^2k_j^3+\frac{1}{8}\sum_{i}k_i^3 \right]  
\label{Ac}
\eea
and where $K=k_1+k_2+k_3$.
\begin{figure}[h]
  \begin{center}
    \subfloat[Shapes of $S^{\lambda}$ in Eq.~\refeq{Alam} (left) and $S^{c}$ in Eq.~\refeq{Ac} (right).]{
      \includegraphics[width=0.8\textwidth]{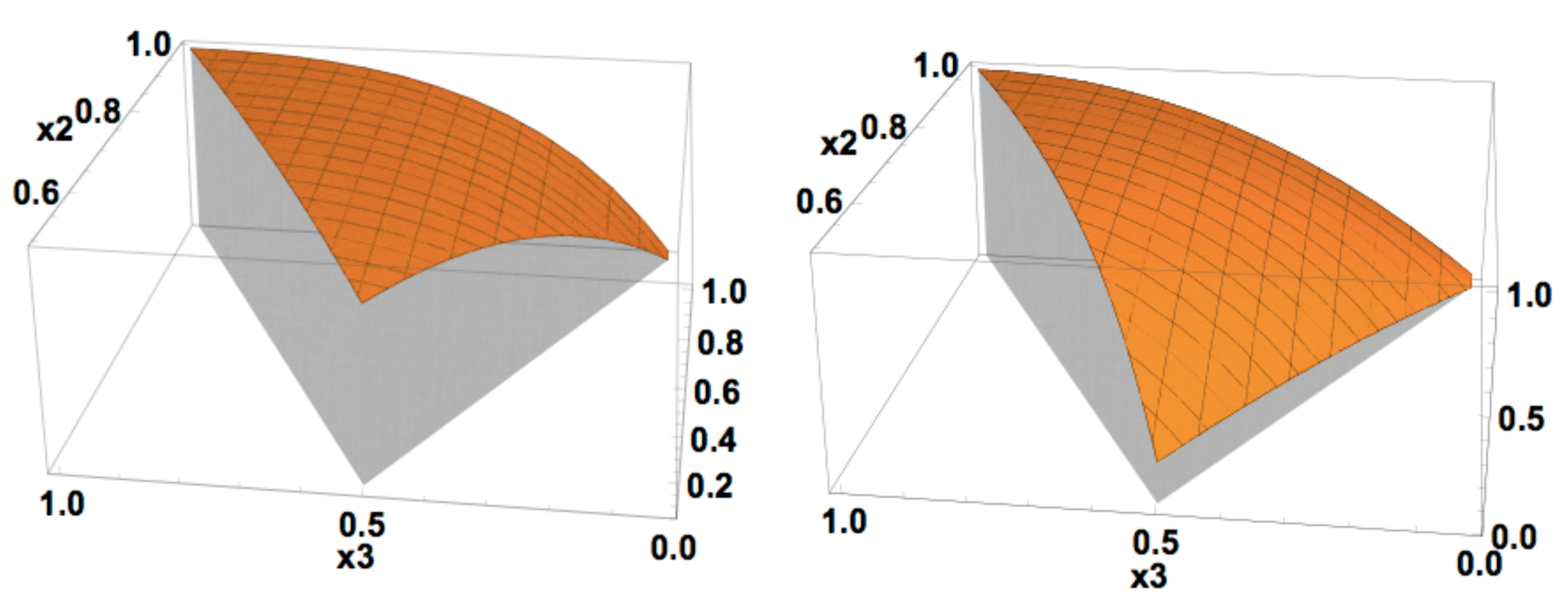}
      \label{fig:subfig1}
                         }
                         \\
    \subfloat[Shapes of $S^{{\rm eq}}$ in Eq.~\refeq{ansatz_eq} (left) and of the absolute value of $S^{{\rm orth}}$ in Eq.~\refeq{ansatz_orth} (right).]{
      \includegraphics[width=0.8\textwidth]{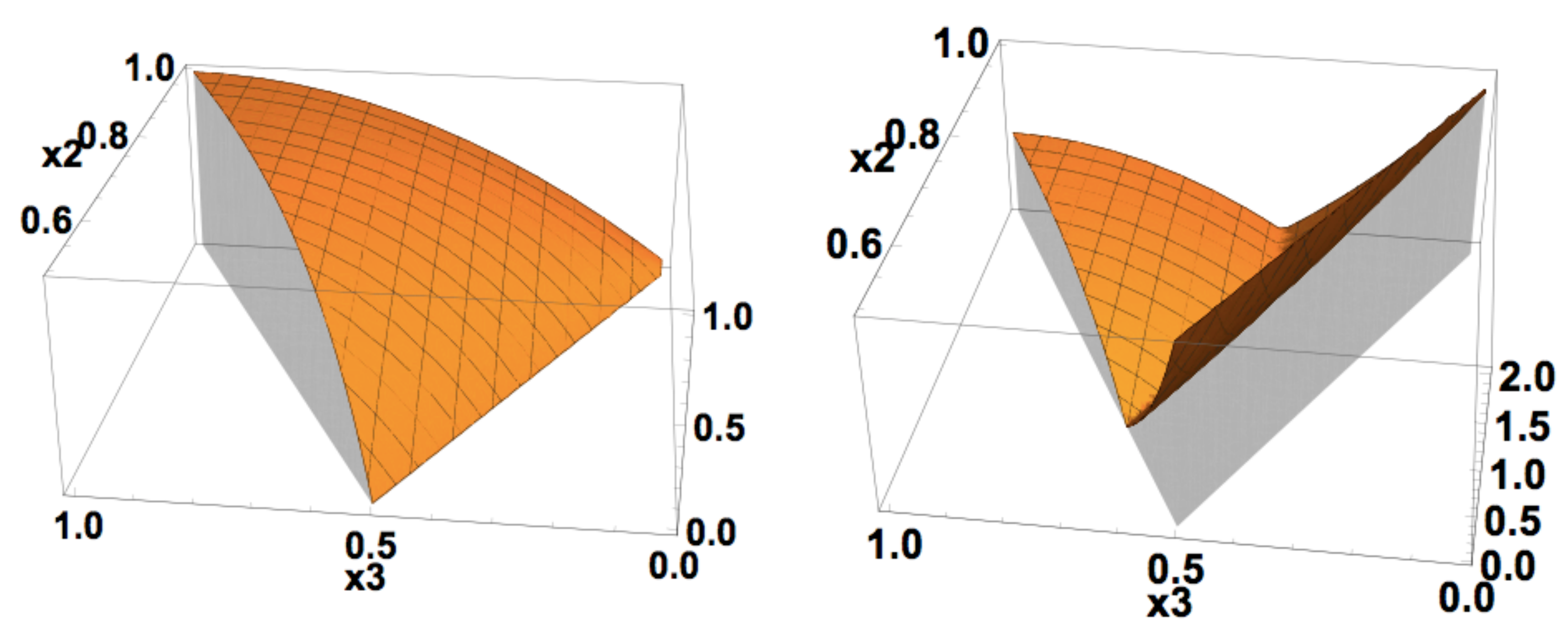}
      \label{fig:subfig2}
                         }
    \caption{Different shapes $S^X\left(1,x_2,x_3\right)$ as a function of $\left(x_2,x_3\right)$. We set them to zero outside the region $1-x_2 \leq x_3 \leq x_2$ and normalize them to one for equilateral triangles $x_2=x_3=1$.}
    \label{single-field-shapes}
  \end{center}
\end{figure}
One therefore finds $f_{{\rm NL}}={\cal O} \left(\frac{1}{c_s^2}-1, \frac{\lambda}{\Sigma}\right)$, \textit{i.e.} large non-Gaussianities $f_{{\rm NL}} \gg 1$ are generated when the deviations to the non-canonical structure $P=X-V(\phi)$ are significant. The two shapes $S^\lambda$ and $S^c$ are represented on the top panel of Fig.~\ref{single-field-shapes}, where one can see that they are similar, both peaking in the equilateral limit. In a first approximation, one can represent both of them by a unique simpler template called equilateral \cite{Creminelli:2005hu} (and whose form is motivated by data analysis, see section \ref{observations}):
\bea
&&\hspace*{-4em}S^{\rm eq} = \frac{9}{10} \Feq \left[ -\left(\frac{k_1^2}{k_2k_3} + {\rm 2~perm.}\right)+\left( \frac{k_1}{k_2} + {\rm 5~perm.} \right) -2 \right]\,.
\label{ansatz_eq}
\eea
Although this is not obvious from the cubic action~\refeq{S3-k-inflation}, one can use the linear equation of motion deduced from the second-order action to express $S^{(3)}_{{\rm L.O.}}$ in terms of the interactions $\dot \zeta^3$ and $\dot \zeta (\partial_i \zeta)^2$ only \cite{RenauxPetel:2011sb} (this becomes more transparent in a gauge where the scalar field itself is perturbed, and one can neglect the mixing with gravity for large NG). These derivative interactions are suppressed when any individual mode is far outside the Hubble radius, either by gradient terms, or because of the super-Hubble conservation of $\zeta$. This should not come as a surprise then that the bispectrum is maximal for modes that cross the sound horizon at approximately the same time, \textit{i.e.} for equilateral triangles.

 By looking at Fig.~\ref{single-field-shapes}, one can see that there are small differences between the shapes $S^{\lambda}$ and $S^c$, especially around flattened triangles $k_2+k_3 \simeq k_1$. One can therefore highlight their differences by considering an appropriate linear combination of them that subtracts their similarities. This way, one can cover more efficiently than with $S^{\rm eq}$ alone the two-dimensional space of shapes spanned by $S^{\lambda}$ and $S^c$. A simple template for such a shape, called orthogonal, is given by \cite{Senatore:2009gt}: 
\bea
&&\hspace*{-4em}S^{\rm orth} = \frac{27}{10} \For \left[ -\left(\frac{k_1^2}{k_2k_3} + {\rm 2~perm.}\right)+ \left( \frac{k_1}{k_2} + {\rm 5~perm.} \right) -\frac{8}{3} \right]\,.
\label{ansatz_orth}
\eea
The two shapes $S^{\rm eq}$ and $S^{\rm orth}$ are represented on the bottom panel of Fig.~\ref{single-field-shapes} (we actually plot the absolute value of $S^{\rm orth}$ so that its difference with the equilateral ansatz for flattened triangles is more visible). Although we concentrated on \textit{k}-inflationary scenarios of the type $P(X,\phi)$, it should be clear from the above discussion that these shapes are signatures of derivative interactions in general. They indeed emerge in more general higher-derivative scenarios, such as ghost-inflation \cite{ArkaniHamed:2003uz}, Galileon-like models \cite{Burrage:2010cu,RenauxPetel:2011dv,RenauxPetel:2011uk}, or Horndeski and generalised theories \cite{Gao:2011qe,DeFelice:2013ar,Fasiello:2014aqa}.\\

\noindent {\bf Consistency relations.}--- We have focused above on the characteristic equilateral-type shapes that single-field inflation with derivative interactions can generate. It is also interesting to consider what type of non-Gaussian signal single-field inflation \textit{cannot} generate. In this respect, one can show that, under very general conditions, a phase of inflation driven by a single scalar field generates a bispectrum such that \cite{Creminelli:2004yq}
\be
\lim_{k_3 \to 0} B_\zeta(k_1,k_2,k_3)=(1-n_s(k_1)) P_\zeta(k_1) P_\zeta(k_3)\,
\label{consistency}
\ee
where 
\be
n_s(k)-1=\frac{\d \,{\rm ln} \left[ k^3 P_\zeta(k) \right]}{\d\, {\rm ln}\, k}
\ee
is the scalar spectral index. In other words, the squeezed limit of the bispectrum is suppressed by $1-n_s$ and vanishes for a perfectly scale-invariant power-spectrum. Given the observed smallness of the deviation from scale-invariance (see Eq.~\refeq{ns}), this implies that a detection of a large bispectrum signal in the squeezed limit ($f_{{\rm NL}}^{{\rm sq}} \gtrsim 1$) would rule out all models of inflation based on a single scalar field, irrespective of any details like the form of the potential or the kinetic term! Given the importance of this relation, let us give a sketch of its proof. It relies on the fact that a super-Hubble conserved curvature perturbation locally acts as a background field, only rescaling the spatial coordinates within a given Hubble patch (see Eq.~\refeq{gij}). By writing $\langle \zeta_{\bkone}  \zeta_{\bktwo}  \zeta_{\bkthree} \rangle= \langle \langle \zeta_{\bkone}  \zeta_{\bktwo} \rangle_{\zeta_{\bkthree}} \zeta_{\bkthree}  \rangle$, where $\langle \zeta_{\bkone}  \zeta_{\bktwo} \rangle_{\zeta_{\bkthree}}$ is the expectation value of $\zeta_{\bkone}  \zeta_{\bktwo}$ given that $\zeta_{\bkthree}$ has a particular value, one can see that  the squeezed limit bispectrum ($k_3 \to 0$) measures the effect of a long-wavelength fluctuation on the short wavelength power spectrum. From the argument above, this effect is proportional to the change of the power-spectrum under a rescaling of the spatial coordinates, or equivalently, to its deviation from perfect scale-invariance. 

As should be clear from this sketch, its crucial argument is the constancy of $\zeta$ on super-Hubble scales. In this respect, it should be said that, strictly speaking, single-field inflation only implies the existence of a super-Hubble constant mode for $\zeta$, its other mode usually decaying exponentially. However, by choosing ad hoc potentials and initial conditions, it is possible to render this decaying mode significant on super-Hubble scales, thus violating the consistency relation (see \textit{e.g.} \cite{Chen:2013aj,Mooij:2015yka}). This mild limitation does not however limit the importance of the relation \refeq{consistency}, which has been derived using various symmetry arguments, as well as generalised to higher-order correlation functions and slightly different situations (see \textit{e.g.} Refs.~\cite{Creminelli:2012ed,Senatore:2012wy,Hinterbichler:2012nm,Assassi:2012zq,Goldberger:2013rsa,Hinterbichler:2013dpa,Berezhiani:2013ewa,Pimentel:2013gza}).

\section{Multi-field models}

In this section, we consider models characterised by the presence of several light scalar degrees of freedom during inflation. A prototypical exemple is the one of a collection of canonical scalar fields $\phi^I$ minimally coupled to gravity and interacting through a generic potential $V(\phi^I)$. Contrary to the single field case, one can not remove here all the fluctuations of the scalar fields by a suitable choice of coordinates. Instead, it is convenient in that case to use the so-called spatially flat gauge, in which the spatial part of the metric $g_{ij}$ takes its background form and the field fluctuations $Q^I$ are made explicit. The second-order action then takes the form \cite{Sasaki:1995aw,GrootNibbelink:2001qt,Seery:2005gb,Langlois:2008mn}
 \begin{eqnarray}
&&\hspace*{-2em}S^{(2)}= \int  \d t\, \d^3x \,a^3\left[ \sum_I \left( (\dot Q^I)^2-\frac{(\partial_i Q^I)^2}{a^2} \right)
-
\underbrace{
\left( V_{,IJ} -\frac{1}{a^3} \Tdot{\left[\frac{a^3}{H} \dot \phi_
I \dot \phi_J\right]} \right)
}_{M_{IJ}}  Q^J Q^J \right]\,.
\label{S2-multifield}
\end{eqnarray}
When the effective mass matrix $M_{IJ}$ is negligible compared to $H^2$ around Hubble crossing, \textit{i.e.} in the case of effectively light degrees of freedom, it follows from Eq.~\refeq{S2-multifield} that the fluctuations $Q^I$ acquire independent almost scale-invariant power spectra soon after Hubble crossing:
\be
C^{AB}(k)=\frac{H_*^2}{2 k^3} \delta^{AB}\,. \label{CAB-result}
\ee
At this stage, $N$-field inflation resembles $N$ copies of single-field inflation. The crucial difference lies in the super-Hubble regime. One can indeed show that, for any relativistic theory of gravity and any matter content, the local energy conservation implies that, for linear perturbations about an FLRW metric \cite{Wands:2000dp}:
\be
\dot \zeta=- H \frac{\delta p_{{\rm nad}}}{\rho+p} +{\cal O} \left(\frac{k}{a H} \right)^2\,
\label{zeta-dot}
\ee
where $\rho$ and $p$ are the matter energy density and pressure respectively, and $\delta p_{{\rm nad}}$ is the non-adiabatic pressure perturbation, defined as 
\be
\delta p_{{\rm nad}}=\delta p-\frac{\dot p}{\dot \rho} \delta \rho\,.
\ee
In (attractor) single-field models, all variables $x$ share the same time shift, $\delta t = \frac{\delta x}{\dot x}$, along a \textit{single} phase-space trajectory. As a consequence, the so-called entropy perturbation between any two quantities, $\Gamma_{x y} \equiv \frac{\delta x}{\dot x}-\frac{\delta y}{\dot y}$, vanishes, and so does the non-adiabatic pressure perturbation on large scales in particular. On the contrary, in a multifield model, there is a family of inflationary trajectories, and perturbations in directions orthogonal to the classical path in field space lead to cosmological evolutions that are not simply translations of it, sourcing the non-adiabatic pressure perturbation, and hence the curvature perturbation $\zeta$ on super-Hubble scales \cite{Starobinsky-Yokoyama}. 
 \begin{figure}
 \centering
\includegraphics*[width=7.5cm]{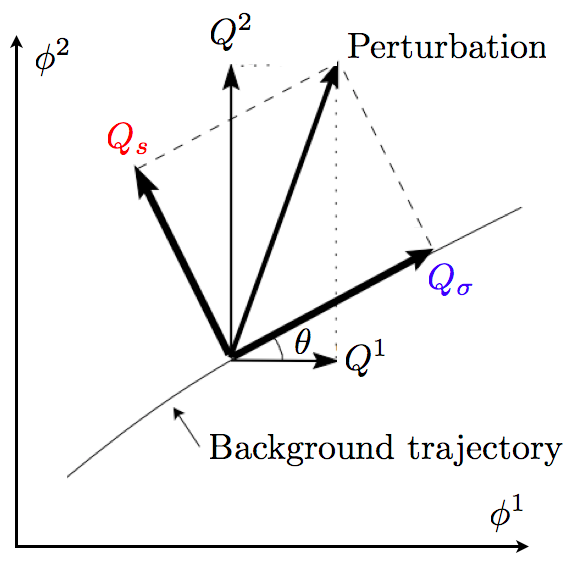}
\caption{In multifield inflation, fluctuations are conveniently decomposed into an instantaneous adiabatic mode, along the background trajectory, and instantaneous entropic/isocurvature modes, orthogonal to it \cite{Gordon:2000hv}.
\label{Adiabatic-entropic}}
\end{figure}

A simple example is provided by two-field inflation. One can show in this context that $\delta p_{{\rm nad}}$ is proportional on super-Hubble scales to $\Gamma_{\phi^1 \phi^2}=\frac{Q^1}{\dot \phi^1}-\frac{Q^2}{\dot \phi^2}$.
This quantity is proportional to the projection of the perturbations in the direction orthogonal to the background trajectory: $Q_s=-\sin(\theta) Q^1+\cos(\theta) Q^2$, where one has introduced the angle $\theta$ such that $\tan(\theta)=\dot \phi^2/ \dot \phi^1$ (see Fig.~\ref{Adiabatic-entropic}). This type of perturbation is called entropic, or isocurvature. On the contrary, one can show that the curvature perturbation is given on large scales by $\zeta=-\frac{H}{\sqrt{(\dot \phi^1)^2+(\dot \phi^2)^2}} Q_{\sigma}$, where \\ $Q_\sigma=\cos(\theta) Q^1+\sin(\theta) Q^2$, called the adiabatic perturbation, is the projection of the perturbations along the direction of the background trajectory. In details, the equation \refeq{zeta-dot} reads, in this context \cite{Gordon:2000hv}:
\be
\dot \zeta =-2\frac{H \dot \theta}{\sqrt{(\dot \phi^1)^2+(\dot \phi^2)^2}} \, Q_s+{\cal O} \left(\frac{k}{a H} \right)^2\,.
\ee
In other words, the isocurvature perturbation sources the adiabatic curvature perturbation on super-Hubble scales as soon as the trajectory bends in field space, \textit{i.e.} $\dot \theta \neq 0$. This genuine multifield effect can drastically modify the properties of the cosmological perturbations and should be accounted for in any model with multiple degrees of freedom. \\

\begin{figure}
 \centering
\includegraphics*[width=8.5cm]{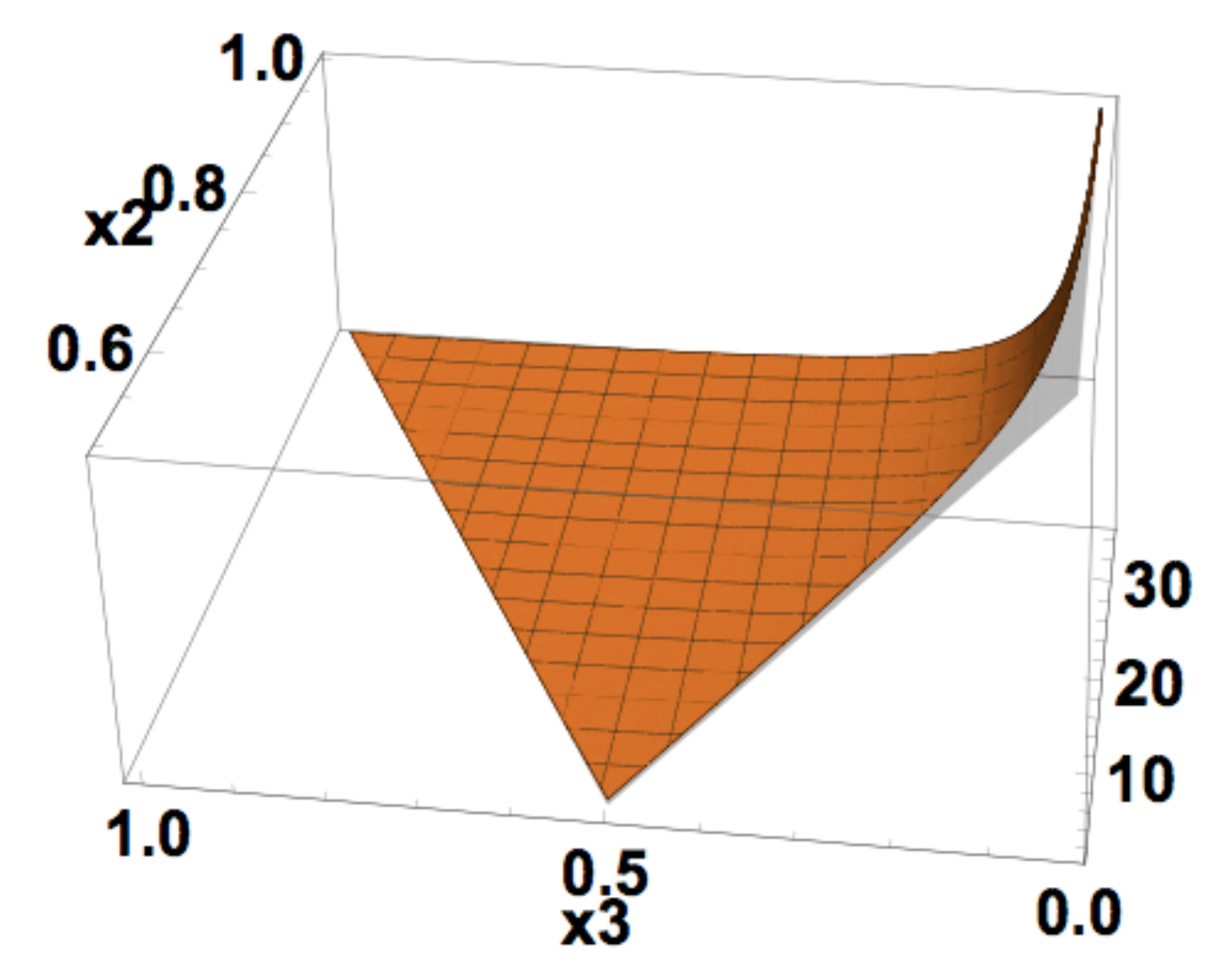}
\caption{Shape of $S^{\rm loc}$ in Eq.~\refeq{ansatz_loc}, with the same specifications as in Fig.~\ref{single-field-shapes}.
\label{local-shape}}
\end{figure}

\noindent {\bf Local non-Gaussianities.}--- Our discussion here was at the linear level up to now but can be extended to non-linear orders (see \textit{e.g.} Refs.~\cite{Langlois:2005ii,Langlois:2006vv,Lehners:2009ja}): NG of the curvature perturbation can be generated by the non-linearities of the transfer mechanism from the super-Hubble isocurvature perturbations, or can be linearly inherited from intrinsic non-Gaussianities in the isocurvature sector, which is less constrained than the adiabatic one by the slow-roll requirements. We used the bending of the trajectory in multifield inflation as a representative concrete example of this transfer mechanism \cite{Bartolo:2001cw,Bernardeau:2002jy,Bernardeau:2002jf,Rigopoulos:2005xx,Rigopoulos:2005ae,Vernizzi:2006ve,RenauxPetel:2008gi}, but it can arise at the end of inflation \cite{Dvali:2003em,Kofman:2003nx,Dvali:2003ar,Bernardeau:2004zz}, or even after, like in the curvaton scenario \cite{Linde:1996gt,Enqvist:2001zp,Lyth:2001nq,Moroi:2001ct,Lyth:2002my}. However, whatever the specific model that is considered, the shape of the associated non-Gaussianities takes a universal form. Indeed, these processes arise on super-Hubble scales, where gradients can be neglected by definition, so that the physics become local in real space, and hence non-local in Fourier space, \textit{i.e.} the bispectrum correlates large and small scale Fourier modes. This becomes transparent in the language of the $\delta N$ formalism presented in section \ref{Methods}, where these scenarios correspond to large super-Hubble classical non-linearities, \textit{i.e.} to the second contribution to the bispectrum in Eq.~\refeq{zetabispectrum}. With the result \refeq{CAB-result} for the spectra of light scalar fields, 
this translates into the so-called local shape \cite{Komatsu:2001rj,Gangui,Verde:1999ij} (see Fig.~\ref{local-shape})
 \bea
S^{\rm loc} = \frac{3}{10} \Floc
\left( \frac{k_1^2}{k_2k_3} + {\rm 2~perm.} \right) 
\label{ansatz_loc}
\eea
where 
\be
\Floc \equiv  \frac{5}{6} \frac{N_A N_B N^{AB}}{\left(N_C N^C\right)^2} \,.
  \label{fNLloc}
 \ee
Large non-Gaussianities can hence be generated in the squeezed limit $k_3 \ll k_2 \simeq k_1$, where the shape \refeq{ansatz_loc} peaks, in sharp contrast with single-field models, in which the consistency relation \refeq{consistency} severely bounds the amplitude of the non-Gaussian signal in this limit.

A similar discussion holds at higher orders in perturbation theory. In particular, the last two contributions in Eq.~\refeq{trispectrum-general} generate in these models a trispectrum of the form
\bea
&&\hspace*{-2.2em} T_\zeta^{\rm loc} (\bk_i)=
 \tau_{{\rm NL}}\left[P_\zeta(k_{13})P_\zeta(k_3)P_\zeta(k_4) +\rm{perm.}\right] +\frac{54}{25}g_{{\rm NL}}\left[P_\zeta(k_2)P_\zeta(k_3)P_\zeta(k_4)+\rm{perm.}\right]
\label{tauNLgNLdefn} 
\eea
with
\be
  \tau_{{\rm NL}}=\frac{N_{AB}N^{AC}N^BN_C}{(N_DN^D)^3}\, 
   \label{tauNLmultifield}
  \ee
and
  \be
  g_{{\rm NL}}= \frac{25}{54}\frac{N_{ABC}N^AN^BN^C}{(N_DN^D)^3}\,.
   \label{gNLmultifield}
 \ee
 Similarly to the case of the local bispectrum, the $\tau_{{\rm NL}}$ and $g_{{\rm NL}}$ shapes
peak in particular `soft' limits of the tetrahedron, with respectively a diagonal (internal wavevector) and one side of the general wavevectors being much smaller than the others.
The parameter $g_{{\rm NL}}$ corresponds to a contact interaction and is related to intrinsic cubic non-linearities (the third term in Eq.~\refeq{dev-deltaN}), while $\tau_{{\rm NL}}$ corresponds to a scalar-exchange interaction and is related to second-order non-linearities, like $\Floc$. The two are actually related by the inequality \cite{Suyama:2007bg}
\be
\tau_{NL} \ge \left(\frac{6}{5}\Floc \right)^2\,,
\label{tauNL-fNL}
\ee
which is a particular case of a more general consistency relation \cite{Assassi:2012zq}. The bound \refeq{tauNL-fNL} is saturated in scenarios where a single field beyond the inflaton generates the curvature perturbation, and larger values indicate a more complicated dynamics.

\section{Beyond the simplest non-Gaussian shapes}

The non-Gaussian shapes we have described until now --- local, equilateral, orthogonal --- are well motivated and cover interesting broad classes of inflationary models. However, they only constitute the tip of the non-Gaussian iceberg. In this section, we briefly mention some other non-Gaussian signatures.\\

\noindent {\bf Excited initial states.}--- In models with excited (\textit{i.e.} non Bunch-Davies) initial states, the usual mode function \refeq{mode-function} in $e^{-i k \tau}$ acquires a small negative energy component $\sim e^{i k \tau}$. This results in (so far unobserved) oscillations in the primordial power spectrum \cite{Martin:2000xs,Danielsson:2002kx,Greene:2005aj}, which severely constrains the magnitude of the effect. As for the bispectrum \refeq{equ:first}, its computation involves an integration of the product of three mode functions with wavenumbers $k_1, k_2$ and $k_3$. Hence the main correction to the Bunch-Davies result is to replace one of the $k_i$'s with $-k_i$ in this calculation, turning the usual factors of $1/K$ coming from the integration of $e^{-i K \tau}$
into $1/(k_2+k_3-k_1)$ and its permutations. As a result, although detailed predictions are model-dependent (see \textit{e.g.} \cite{Chen:2006nt,Holman:2007na,Meerburg:2009ys}), the most important feature of this type of modification is to enhance the bispectrum for flattened triangles, with $k_2+k_3 \simeq k_1$. A simple ansatz that captures this main characteristic is given by \cite{Meerburg:2009ys}
 \bea
&&\hspace*{-4em}S^{\rm flat} = \frac{9}{10} f_{{\rm NL}}^{{\rm flat}} \left[ \left(\frac{k_1^2}{k_2k_3} + {\rm 2~perm.}\right)- \left( \frac{k_1}{k_2} + {\rm 5~perm.} \right) +3 \right]\,,
\label{ansatz_flat}
\eea
which is simply a linear combination of the equilateral and orthogonal templates.\\

\noindent {\bf Scale-dependent resonance and feature models.}--- Periodic features in the inflationary potential naturally emerge when trying to UV-complete large field inflationary models, in particular in a string theory context \cite{Flauger:2009ab,Hannestad:2009yx}. These features can resonate with the sub-Hubble oscillations of the inflationary perturbations, generating an oscillatory bispectrum whose shape takes the form \cite{Chen:2008wn,Chen:2010bka}
\be
S^{{\rm res}}=\frac{9}{10} f_{{\rm NL}}^{{\rm res}} \,{\rm sin} \left[ \,C\, {\rm ln} \left(k_1+k_2+k_3 \right) + \phi \,\right]\,
\label{res}
\ee
in the simplest model, where $C$ depends on the periodicity of the features.

Temporary violations of the slow-varying evolution can occur if there are sharp features in the inflationary potential or sharp turns in field space in multifield inflation for example \cite{Chen:2006xjb,Achucarro:2010da,Adshead:2011jq,Achucarro:2012fd}. This can generate oscillatory correlation functions of a type different from the one in resonance models, the simplest bispectrum signal being of the form
\be
S^{{\rm feat}}=\frac{9}{10} f_{{\rm NL}}^{{\rm feat}} \,{\rm sin} \left[\, \omega \,\left(k_1+k_2+k_3 \right) + \phi\, \right]\,,
\label{feat}
\ee
where $\omega$ again depends on the specific disruption of the attractor regime. Contrary to the simple almost scale-invariant shapes discussed in the previous sections, the two templates \refeq{res} and \refeq{feat} provide examples of well motivated shapes whose scale-dependence is non-trivial.\\

\noindent {\bf Quasi-single-field inflation.}--- Technically natural models in supersymmetry often have extra fields beyond the inflaton with masses $m$ of order the Hubble scale during inflation \cite{Chen:2009we,Chen:2009zp,Baumann:2011nk,McAllister:2012am}.
In this sense, they interpolate between truly multifield inflation, where all fields are almost massless ($m/H \ll 1$), and single field inflation, which can be thought of as the limit $m/H \to \infty$. In these models, the conversion of non-linearities from the iscocurvature sector to the curvature one results in a bispectrum shape 
\be
S_{{\rm QSI}}=\frac{3^{\frac92-3 \nu}}{10} f_{{\rm NL}}^{{\rm QSI}} \frac{(k_1^2+k_2^2+k_3^2)(k_1 k_2 k_3)^{\frac12 -\nu}}{(k_1+k_2+k_3)^{\frac72-3 \nu}}
\ee
that behaves in the squeezed limit as $S_{{\rm QSI}} \propto \left(\frac{k_3}{k_1}\right)^{\frac12-\nu}$, where $\nu =\sqrt{\frac94-\frac{m^2}{H^2}}$. This scaling behavior is intermediate between that of the local shape $\left( k_1/k_3 \right)$ and that of the equilateral shape $\left( k_3/k_1 \right)$, exemplifying how soft limits can act as a particle detector.\\

\noindent {\bf The trispectrum beyond the local shapes.}--- Beyond the local trispectrum \refeq{tauNLgNLdefn} that arises in multifield inflation, the intrinsic quantum trispectrum (the first line in Eq.~\refeq{trispectrum-general}) can be large in models with derivative interactions for instance \cite{Chen:2009bc,Arroja:2009pd,Senatore:2010jy,Bartolo:2010di,Senatore:2010wk,Renaux-Petel:2013ppa}, with shapes that are counterparts of the equilateral and orthogonal bispectra \cite{Renaux-Petel:2013wya}. Additionally, classical and quantum effects combine in multifield models with derivative interactions (see the second line in Eq.~\refeq{trispectrum-general}). Some of these models can hence generate a superposition of the local and equilateral bispectra, with a particular shape of the trispectrum whose amplitude is given by the product $\Feq \Floc$ \cite{RenauxPetel:2009sj}.\\

\noindent {\bf Inflation with gauge fields.}---  A coupling of the inflaton to the kinetic term of a gauge field $A^{\mu}$, of the type ${\cal L} \supset -I^2(\phi) F_{\mu \nu} F^{\mu \nu}$, can generate a bispectrum enhanced in the squeezed limit, with a specific signature in the form of a non-trivial dependence on the angle between the small and large wavevectors \cite{Barnaby:2012tk,Bartolo:2012sd,Shiraishi:2013vja}.\\
 Large field models with an approximate shift symmetry also naturally contain 
a coupling between a pseudoscalar axion inflaton and a gauge field ${\cal L} \supset -(\alpha/4f) \phi F_{\mu \nu} \tilde F^{\mu \nu}$, where $\alpha$ is a dimensionless parameter and $f$ is the axion decay constant. In these scenarios, gauge field quanta are produced by the background evolution of the inflaton, and in turn feed the curvature perturbation through an inverse decay process of the gauge field, resulting in a bispectrum shape strongly correlated with the equilateral template \cite{Barnaby:2010vf,Barnaby:2011vw,Barnaby:2011qe,Meerburg:2012id,Linde:2012bt}.\\

\noindent {\bf Isocurvature non-Gaussianities.}--- Multi-field inflationary models can generate residual primordial isocurvature perturbations after inflation. Although their impacts on primordial power spectra are tightly constrained \cite{Ade:2015lrj}, they can contribute significantly at the level of the bispectrum, producing in general  both a pure isocurvature bispectrum and mixed bispectra because of their correlation with the adiabatic perturbation \cite{Langlois:2011hn,Langlois:2012tm}.

\section{Observational probes and constraints} 
\label{observations}

Non-Gaussianities can be constrained by studying the signatures they leave in the anisotropies of the CMB and in the large scale mass distribution in the Universe. Amongst the two, the CMB provides our cleanest window into 
the physics of the primordial universe,
 because its anisotropies are to a very good approximation linearly related to the primordial fluctuations. On the contrary, structure formation is an intrinsically non-linear process, and thus offers a more convoluted probe of 
the initial conditions. The CMB data, and in particular recent ones from the \textit{Planck} mission, provide us with the best constraints so far \cite{Ade:2013ydc,Ade:2015ava}, and we briefly explain in the following how they have been derived (see the contribution by Bouchet to this volume for more details). We then turn more qualitatively to LSS, from where drastic improvements should originate in the future \cite{Alvarez:2014vva}.\\

\begin{figure}
 \centering
\includegraphics*[width=10.5cm]{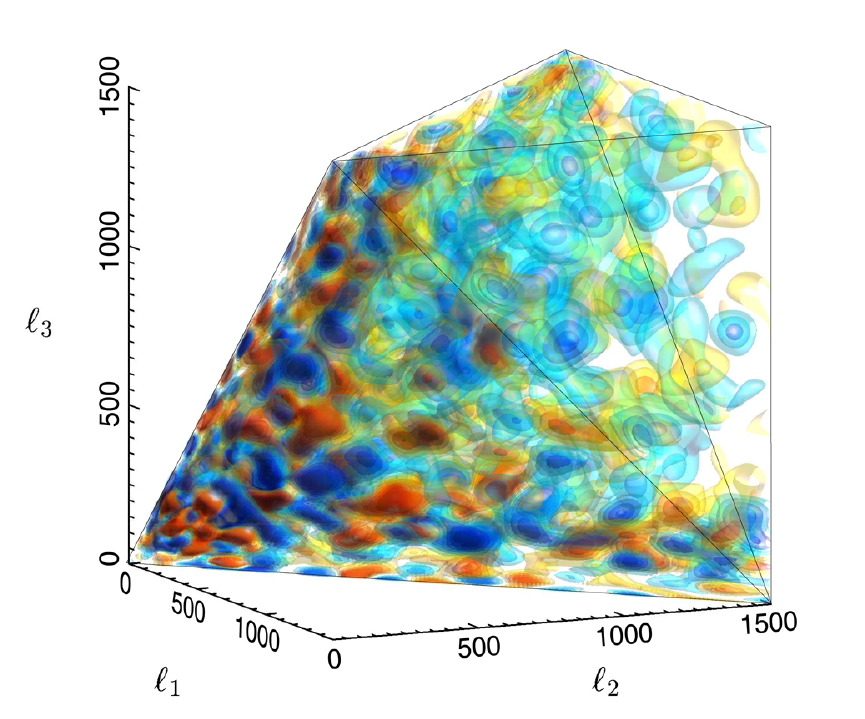}
\caption{\textit{Planck} modal reconstruction of the CMB temperature reduced bispectrum, plotted using several density contours \cite{Ade:2015ava}.
 \label{CMB-bispectrum}}
\end{figure}

\noindent {\bf Primordial non-Gaussianities in the CMB and \textit{Planck} constraints.}--- The CMB maps of temperature and polarisation anisotropies are analysed by means of a decomposition into spherical harmonics:
\bea
\frac{\Delta T}{T}({\boldsymbol {\hat n}})&=&\sum_{\ell m} a^T_{\ell m} Y_{\ell m}({\boldsymbol {\hat n}}) \\
E({\boldsymbol {\hat n}})&=&\sum_{\ell m} a^E_{\ell m} Y_{\ell m}({\boldsymbol {\hat n}}) \,
\eea
where the link between the CMB multipoles $a_{\ell m}^X$ ($X={T,E}$) and the primordial curvature perturbation $\zeta$ is well known at linear order (see the contribution by Durrer to this volume). Using statistical rotational invariance, the CMB angular bispectrum, \textit{i.e.} the three-point correlator of the $a_{\ell m}$'s, can be written as
\be
\langle a_{\ell_1 m_1}^{X_1} a_{\ell_2 m_2}^{X_2} a_{\ell_3 m_3}^{X_3} \rangle ={\cal G}^{\ell_1 \ell_2 \ell_3}_{m_1 m_2 m_3} b_{\ell_1 \ell_2 \ell_3}^{X_1 X_2 X_3}\,
\ee
where ${\cal G}^{\ell_1 \ell_2 \ell_3}_{m_1 m_2 m_3}$ is a known geometrical factor related to Wigner $3 j$-symbols, and $b_{\ell_1 \ell_2 \ell_3}^{X_1 X_2 X_3}$ is referred to as the reduced bispectrum. It is the counterpart in $\ell$ space of the primordial shape $S$ in Eq.~\refeq{def-S}, to which it is linearly related. Importantly, the signal one is looking for is way to small for a mode-by-mode measurement of the (reduced) bispectrum. Instead, one compares theoretically motivated bispectrum templates to the observed one, and fit for their overall amplitudes ($\Feq, \Floc$ ...). The corresponding optimal estimator is well known \cite{Creminelli:2005hu}, but its implementation is very challenging. The computational cost of its direct evaluation is totally prohibitive for high-resolution data like the ones of \textit{Planck}, and several approaches have been developed to circumvent this problem. A central idea is to use templates that are separable, \textit{i.e.} written as linear combinations of products of functions of respectively $k_1, k_2, k_3$, which renders the problem tractable \cite{Komatsu:2003iq}. The local shape \refeq{ansatz_loc} is of this type, while the equilateral and orthogonal ones (Eqs.~\refeq{ansatz_eq}-\refeq{ansatz_orth}) are separable approximations of original non-separable shapes, and were devised with this data-analysis perspective in mind. Building on this idea, modal estimators are based on decomposing the bispectrum (both the theoretical ones and the observed one) into a sum of uncorrelated separable templates forming a complete basis in bispectrum space, and measuring the amplitude of each \cite{Fergusson:2009nv,Fergusson:2010dm}. This way, one can reconstruct once and for all the total bispectrum (see Fig.~\ref{CMB-bispectrum}) and compare it to any theoretical shape, separable or not.
At last, by exploiting the fact that angular bispectra of interest are generally smooth functions in harmonic space, binned estimators, which do not rely on separability, have also been developed and provide a complementary approach \cite{Bucher:2009nm}. 

Based on these methods, the \textit{Planck} collaboration performed a comprehensive analysis of their data, multiplying tests and cross-validations, quoting as their result for the three main shapes:
\be
\left(
\begin{array}{ll}
\Floc=2.5 \,\pm\, 5.7 \\
\Feq=-16\,\pm \,70 \\
 \hspace{-0.16cm} \For=-34\, \pm \,33
\end{array} \right) (68 \,\% \,{\rm CL}, T)
\label{constraints-T}
\ee
from temperature data alone, and
\be
 \hspace{+0.32cm} \left(
\begin{array}{ll}
\Floc=0.8\, \pm \, 5.0 \\
\Feq=-4 \,\pm \, 43 \\
 \hspace{-0.16cm} \For=-26 \, \pm \, 21
\end{array} \right) (68 \,\% \,{\rm CL}, T+E)
\label{constraints-T+E}
\ee
from combined temperature and polarisation data. Despite having passed several consistency and robustness tests, it should be said that the polarisation data are known to suffer from some systematics and the $(T+E)$ constraints should hence be considered as \textit{preliminary}.

Besides these estimates, the \textit{Planck} collaboration constrained an impressively wide range of NG bispectra, including all the ones mentioned in this review and more. While they found no evidence for primordial non-Gaussianities, the `hints' of NG reported in the \textit{Planck} 2013 analysis of oscillatory patterns are confirmed and reinforced with a more rigorous statistical analysis. To be precise, there appears to be no evidence for any individual feature or resonance model at a particular frequency, but multiple feature or resonance models might explain the apparently high non-Gaussian signal observed in the bispectrum reconstruction (see Fig.~\ref{CMB-bispectrum}). The evidence of several of these models increased to more than $3 \sigma$ with the additional polarisation information, taking into account the `look-elsewhere' effect, \textit{i.e.} the fact that when scanning across the parameter space of models, Gaussian noise can lead by chance to a large apparent signal. No strong claim can be made at this stage though, as any information gathered from the polarisation data can only be considered as preliminary.

At the level of the trispectrum, besides the constraint $\tau_{{\rm NL}} < 2800 \,(95\,\%\, {\rm CL}) $ from the \textit{Planck} 2013 analysis \cite{Ade:2013ydc}, the \textit{Planck} 2014 analysis further gives $g_{{\rm NL}}=(-9.0\pm 7.7) \times 10^4$, as well as estimates of two intrinsically quantum trispectra.

Although the study of the angular bispectrum and trispectrum provides optimal constraints, let us point out that very good and consistent constraints on $\Floc$ and $g_{{\rm NL}}$ were also obtained by using completely different tools, known as Minkowski Functionals, and which are based on the study of morphological features of random fields \cite{Ducout:2012it}.

Eventually, one word about non-primordial NG (see the contribution by Bernardeau to this volume for further details): the high-quality data of \textit{Planck} are such that, despite the small amplitude of primordial fluctuations, nonlinear effects from General Relativity itself, which are present even for perfectly Gaussian initial conditions, have to be taken into account. Most notably, the correlation between the gravitational lensing of the CMB anisotropies and the so-called integrated Sachs-Wolfe effect gives rise to a secondary CMB bispectrum that bias the naive estimate of $\Floc$ by about 5. This contribution, of the same order as the statistical error in \refeq{constraints-T}-\refeq{constraints-T+E}, has been subtracted from the results.\\

\noindent {\bf Primordial non-Gaussianities in LSS.}--- An obvious way to probe primordial non-Gaussianities in LSS is via the study of higher-order correlators of the density field of galaxies, like the halo bispectrum. However, extracting the primordial information from such measurements is hampered by the significant non-linearities caused by the gravitational clustering itself. Hence, despite recent progress, we do not currently understand the galaxy density contrast to the accuracy required by the galaxy surveys, and there is no constraint to date on primordial NG from the halo bispectrum. 

There have been impressive progress in another direction though, called the scale-dependent bias. LSS surveys do not directly observe the mass distribution in the universe, but rather luminous biased tracers of it like the distribution of galaxies. In the presence of local type NG, it has been discovered recently that the bias between the density contrast of tracers and the one of matter acquires a non-trivial scale-dependence behaving like $1/k^2$ at large scales, making it a very sensitive probe \cite{Dalal:2007cu}. Intuitively, NG in the squeezed limit implies
that long-wavelength fluctuations in the gravitational potential (equivalently the curvature perturbation) locally rescale the amplitude of small scale matter fluctuations, thus affecting the threshold needed to form (halos of) galaxies, and thus the bias, the $1/k^2$ effect coming from the Poisson equation relating the Laplacian of the gravitational potential to the matter over-density. This effect manifests itself at the level of the halo \textit{power spectrum}, which greatly simplifies the analysis compared to higher-order correlators. Using photometric quasars data from SDSS, the most stringent constraints to date give \cite{Leistedt:2014zqa} $-49 < \Floc < 31$ and $-2.7 \times 10^5< g_{{\rm NL}} < 1.9 \times 10^5$ when constrained individually, and $-105 < \Floc < 72$ and $-4.0 \times 10^5< g_{{\rm NL}} < 4.9 \times 10^5$ in a joint analysis (all at 95\,\%\, {\rm CL}). These constraints are weaker than recent ones from the CMB, but provide a useful complementary and consistent picture.
   
\section{Implications for early Universe physics and perspectives}

The simplest single-field slow-roll models of inflation recently passed very stringent tests due to the lack of measurable deviations from Gaussianity of the primordial fluctuations. The constraints that have been obtained strongly limit the different alternative mechanisms that have been proposed to explain the seeds of cosmological perturbations (see Ref.~\cite{Ade:2015ava} for detailed constraints on numerous specific scenarios). 

The bounds on equilateral and orthogonal NG
translate into a limit on the speed of sound of general single-field models of inflation (see section \ref{Single-field}):
\be
c_s \geq 0.020 \quad 95 \,\% \,{\rm CL} \,(T{\rm-only})\,
\ee
and $c_s \geq 0.024 \, (95 \,\% \,{\rm CL})$ with the addition of the preliminary polarisation data. In the language of the effective field theory of inflation \cite{Cheung:2007st,Creminelli:2006xe}, where inflation is studied as the theory of fluctuations of spontaneously broken time translations around a quasi de-Sitter background, the NG parameters can be related to the energy scale of the inflaton self-interactions
as $\zeta f_{{\rm NL}}^{{\rm eq, orth}}  \sim H^2/\Lambda^2$ (see \cite{Baumann:2011su,Assassi:2013gxa,Baumann:2014cja} for details). Using these arguments, one can show that the cosmological dynamics is unaffected by higher-derivative terms for $f_{{\rm NL}}^{{\rm eq, orth}}  \lesssim 1$, corresponding to $c_s \simeq 1$. While current constraints still allow inflation to have wildly different dynamics than slow-roll, reaching the sensitivity $\Delta f_{{\rm NL}}^{{\rm eq, orth}} \simeq 1$ hence represents a well defined theoretical target for future experiments.

From the single-field consistency relation \refeq{consistency}, to local type non-Gaussianities in multifield inflation, through quasi-single-field inflation, we have seen that the behaviour of higher-order correlation functions in squeezed limits is a sensitive probe of the number of degrees of freedom active during inflation. In this respect, the stringent bound \refeq{constraints-T}-\refeq{constraints-T+E} obtained on $\Floc$ severely constrains the dynamics of models in which light degrees of freedom beyond the inflaton play a role on super-Hubble scales. Compared to the bound on $f_{{\rm NL}}^{{\rm eq, orth}}$, it is however harder to interpret it without specific mechanisms in mind, as the link between multifield scenarios and observables is very much model-dependent. 
In the large class of spectator models though, in which inflation is driven by a field while another one generates the primordial fluctuations, $| \Floc | \gtrsim {\cal O}(1)$ is a rather generic prediction \cite{Suyama:2013nva}, like in the curvaton scenario or in modulated reheating. While achieving the sensitivity to test the single-fied consistency relation is currently out of reach, reaching $\Delta \Floc \simeq 1$ therefore provides a challenging but conceivable target of particular interest. 

Due to the limitations set by Silk damping and foregrounds, it is unlikely that the CMB will offer significant improvements in the characterization of the statistical properties of the primordial fluctuations. The impressive constrains on primordial non-Gaussianities from the \textit{Planck} collaboration have indeed essentially achieved optimality, as shown by comparison with Fisher matrix forecasts. The way forward here seems to be LSS, including 21cm line radio surveys \cite{Camera:2015yqa}, which probes complementary smaller scales and enables one to extract genuine 3D information, in contrast with the two-dimensional CMB sphere. Realistic optimised galaxy surveys have the potential to reach $\Delta \Floc \simeq 1$ by using the scale-dependent galaxy bias \cite{dePutter:2014lna}. As for other types of NG,
one will have to resort to higher-order correlators like the halo bispectrum. Although one is hampered by theoretical uncertainties and observational systematics, the theoretical target $\Delta f_{{\rm NL}}^{{\rm eq, orth}} \simeq 1$ is within reach of currently planned surveys \cite{Alvarez:2014vva}. ``When you have exhausted all possibilities, remember this: you haven't.'' supposedly said Thomas Edison.

%
\paragraph{Acknowledgements.} I would like to thank Jean-Philippe Uzan for inviting me to write this review, as well as for his patience for obtaining it. I would also like to thank my colleagues and collaborators for sharing their insights on its material.


\end{document}